# Laser-fabricated porous alumina membranes (LF-PAM) for the preparation of metal nanodot arrays

A. Pereira**, D. Grojo**,  M. Chaker, Ph. Delaporte, D. Guay* and M. Sentis

Second revised version

Smll.200700256

[*]    A. Pereira, M. Chaker, and D. Guay
INRS-EMT
1650 blvd Lionel-Boulet, Varennes (Québec) J3X1S2, Canada
E-mail: guay@emt.inrs.ca

D. Grojo, Ph. Delaporte, and M. Sentis
LP3-UMR6182 CNRS, Université de la Méditerranée
Case 917, 13288 Marseille Cedex 9, France
E-mail : grojo@lp3.univ-mrs.fr

[**]    A. Pereira and D. Grojo contributed equally to this work.

The synthesis of materials with highly ordered structures at the nanometer scale has generated great interest because of their unusual physical properties that allow their use as nanoscale devices in microelectronics, optics, or bio-chemistry. This is especially so of nanodot arrays, where small amounts of materials having a defined shape and size (nanodot) are arranged periodically (array) at the surface of a substrate. However, our ability to fabricate arrays of functional structures with controlled size and shape, on a substrate of choice, using a minimal number of processing steps, is limited and must be improved to exploit the full benefit of such arrays.

Most studies dealing with the preparation of nanodot arrays rely on the use of Porous Alumina Membranes (PAMs), which was pioneered by Martin's group,[1] as evaporation masks.[2-5] These masks are usually prepared by a two-step anodization process of aluminium[6] to generate free-standing nanoporous Anodic Aluminium Oxide (AAO). Since the pore diameter, pore length and pore-pore distance can be varied by choosing the appropriate anodization conditions, AAO membranes are ideal masks for the preparation of nanodot arrays with variable dot size and dot-dot separation distance. However, one drawback of this approach is that it requires numerous chemical steps and, consequently, it is tedious and takes a long time (> 40h). Moreover, once the mask is formed, it must be contacted to the substrate, either through Van der Waals forces and/or mechanical pressing. This step is the most critical one if controllable dot size and shape are to be achieved. Indeed, depending on the quality of the contact, considerable broadening of the deposited structures compared to the PAM pore size is generally observed.[3,4]

Alternatively, an aluminum film can be deposited on a suitable substrate and a porous alumina membrane can be prepared directly on it.[7,8] Using this approach, no lateral spreading of the dots is observed because the PAM is in intimate contact with the underlying substrate. However, this approach imposes restriction on the nature of the substrate and the long processing time remains a constraint.

Attempts to create alternative, high-resolution, and low-cost patterning processes, have also led to the development of several other approaches. Among others, focused ion beam patterning, electron-beam direct writing and nanoimprint lithography have been investigated and proposed as high-resolution top-down techniques to pattern functional materials. However, all of these approaches rely on the use of a resist or polymer process and hence, numerous chemical, thermal, and etching associated steps. Recently also, a nanostenciling approach[9,10] have appeared as a promising method to fabricate nanodot array. This approach drastically reduces the number of processing operations with respect to conventional methods. However, the steps leading to mask fabrication can be cumbersome. Also, as is the case for porous aluminum oxide membrane, the dimension control over the deposited structure is determined by the gap between the stencil and the substrate, which determines the lateral spreading.[9] Thus, the ability to fabricate arrays of functional structures with controlled size and shape, on a substrate of choice, using a minimal number of processing steps, remains an important challenge in nanotechnology.

Several studies have demonstrated the possibility to produce nanoscale features on various substrates by particle-assisted near-field enhancement. To achieve this, a monolayer of microspheres is deposited at the surface of a substrate. When irradiated by a single (short) laser pulse, light tightly concentrates underneath the particles, causing the local melting or the ablation of a small part of the surface.[11-20] Depending on the particle composition and irradiation conditions, the features reported in the literature extend from well-shaped nanocraters[11-14] to nanobumps or cones.[15-18] These structures are indeed local deformation of the substrate, and their composition can not be controlled independently of that of the substrate. This imposes severe limits on the applicability of nanostructured arrays prepared that way.

Herein, we report on an efficient photonic-based method to prepare nanodot array of functional materials, independently of the nature of the substrate. This is achieved by using particle-assisted near field enhancement to first prepare a thin porous aluminium oxide membrane directly at the surface of a substrate. In a second step, this porous aluminium oxide membrane is used as an evaporation mask to prepare nanodot array of functional materials.

The fabrication process is outlined in figure 1. A monolayer of self-assembled spheres is formed onto a thin alumina ($Al_2O_3$) film, which was previously coated on a silicon substrate by means of Pulsed Laser Deposition[21] (Figure 1a). Then, pores are optically drilled in the $Al_2O_3$ film by particle-assisted near field enhancement. This is accomplished through illumination of the spheres with a single nanosecond laser pulse at the wavelength $\lambda_{las}$ = 193 nm. This leads to the local removal of the 20 nm thick $Al_2O_3$ film under each sphere. Since the spheres are arranged in a hexagonal array at the surface of the substrate, the aluminum oxide film is decorated with an ordered arrangement of holes (Figure 1b). Using this Laser Fabricated Porous Alumina Membrane (LF-PAM) as a mask for the deposition of metal (Figure 1c), a series of ordered metal nanodots are formed at the surface of the substrate upon dissolution of the alumina layer (Figure 1d). As we shall see later on, we demonstrated the controllability of this schematic approach by forming an ordered array of gold nanodots.

Numerous studies devoted to nanostructuring by particle-assisted near-field enhancement deal with 1-μm PolyStyrene (PS) spheres. However, Piparia et al.[18] demonstrated recently that, in the near UV region (< 300 nm), the interaction process exhibits a self-limited character. In this spectral region, short wavelengths are strongly absorbed in most polymer materials. Consequently, the initial portion of the laser pulses rapidly warms the PS spheres and induces their thermal-deformation or even their ablation. Under these conditions, the focusing power of the sphere decreases with time, and it was shown that nanobumps previously reported in the literature were actually organic debris coming from the photo-degradation of the PS spheres.[16] To avoid this limitation, we choose to use silica ($SiO_2$) spheres. Also, spheres with radii $R = 250$ nm were chosen to prepare a nanodot array with a larger surface density ($\cong 5 \times 10^8$ dots cm$^{-2}$) than accessible with 1-µm spheres.

The methods available for the preparation of monolayer of self-assembled spheres have recently been reviewed.[22] They include dip-coating, floating on an interface, electrophoretic deposition, physical template-guided self-organization, and spin-coating. Using these techniques, colloidal particles can be arranged into an ordered array through self-organisation. Among the various methods, spin-coating was selected because it is easy to use and does not rely on complex equipments.[22-27] Moreover, as shown elsewhere, the thickness of the particle layer is easily controlled by adjusting the particle loading, the rotation speed and the deposition time. Using spin-coating, defect-free self-assembled single layer of nanospheres with areas ranging from 10 – 100 µm$^2$ are easily prepared.[24,25] More recently, well-ordered arrays of silica nanospheres on 4-inch wafer have been successfully prepared using the same spin-coating technique.[27] It is thus possible to envision the preparation of nanodot array extending on several centimetres.

In this work, monolayer of self-assembled silica spheres ($R = 250$ nm) onto thin alumina films were prepared by spin-coating. A scanning electron microscopy (SEM) micrograph showing the organization of microspheres at the surface of the substrate is shown as Fig. S1 in Supporting Information. As described later on, a nanohole array is created on the alumina film through nanosecond laser illumination of the spheres under appropriate laser fluence. A single laser pulse (2mm x 2mm square laser spot) allows irradiating the entire silica sphere array ($\sim 100$ µm$^2$). Therefore, this rapid and efficient photonic-based method, relying on particle-assisted near field enhancement to synthesize porous alumina membranes can be easily extended to wafer-scale processes. Moreover, the feature scales are controlled by changing the size of the nanospheres and can be reduced to less than 100 nm.

To better understand the focusing power of the silica spheres, figure 2 presents the results of a calculation of the total external near field intensity $|E|^2/I_0$ using the Lorentz-Mie theory. In this calculation, it was assumed that the particles are irradiated with an unpolarized plane wave, and $I_0=|E_0|^2$ is the intensity of the incident beam in the absence of the silica sphere. The intensity maximum is observed just below the rear surface of the sphere with an enhancement factor $|E|^2/I_0 \cong 28$. Finer calculation would have to take into account the presence of the substrate[28,29] and the neighboring particles.[30] However, according to our calculation, the enhancement factor is larger than one over an axial distance that far exceeds the thickness of the Al$_2$O$_3$ film (Figure 2b), and the surface of the substrate is locally illuminated by a bright spot with a Gaussian-like shape and a full width at half maximum of 160 nm (Figure 2c).

The most critical aspect of this approach is the controlled and selective nanodrilling of the alumina film. Al$_2$O$_3$ is a very low absorption material, largely used for optical coatings in the UV region. At 193 nm, the reflectance of Al$_2$O$_3$ is $R \cong 0.09$ and its optical penetration depth is several orders of magnitude larger than the film thickness used in this study.[31] Consequently, the laser light is mainly absorbed by the Si substrate.

We carefully analyzed, by means of Atomic Force Microscopy (AFM) and Scanning Electron Microscopy (SEM), the effect of the pulse energy on the "spheres-Al$_2$O$_3$-Si" stacks. For laser fluence as low as 200 mJ cm$^{-2}$, the silica spheres are ejected from the surface. According to the laser cleaning mechanism identified recently for similar conditions, this is due to the explosive evaporation of the humidity trapped at the interface between the sphere and the aluminum oxide substrate.[32] Also, craters a few nanometers deeps are formed at the surface of the Al$_2$O$_3$ film (AFM observations), showing that this mechanism slightly damages the surface. At laser fluences close to 340 mJ cm$^{-2}$, the silica spheres and the aluminum oxide film are removed. This is evidenced in Figure 3, where relatively well-formed circular patterns are observed in the Al$_2$O$_3$ thin film. The typical diameter of the ablated craters is $130 \pm 15$ nm which is in fair agreement with the laser energy distribution calculated on the basis of the Mie theory (Figure 2). The bottom of the crater is flat, indicating that the underlying Si substrate is not significantly affected by the process. In this situation, the ablation of the Al$_2$O$_3$ layer is due to a spallation mechanism induced by the confinement of the deposited energy at the alumina-silicon interface.[33] For laser fluences above 400 mJ cm$^{-2}$, the Si substrate is damaged. In that regime, sombrero-like structures having a diameter of about 280 nm are observed. The center of these structures protrudes from the original surface of the (undamaged) aluminum oxide film by about $\cong 8$ nm. Recent studies reported similar phenomena at larger size scale with pure silicon substrates[15] The formation of such nanostructures relates to inward flows of the melted materials caused by the inhomogeneous temperature distribution in the heated layer.[18,34]

Figure 4 shows AFM and SEM images of gold nanodot arrays created on Si substrates by performing LF-PAM at intermediate laser fluence (Figure 3). Image analyses allow us to characterize the size and shape of the gold nanostructures. From the object area $S_{dot}$, the average size of the structures $R_{dot} = (S_{dot} / \pi)^{1/2}$ is $48.2 \pm 2.6$ nm ($\pm$ SD). The circularity, given by the ratio of the square of the perimeter to $4\pi S_{dot}$, is $1.47 \pm 0.1$. Each dot has a uniform height (20 nm) and spacing (500 nm) to its neighbor. The height determined by the AFM profile (Figure 4c) is consistent with the original $Al_2O_3$ film thickness. As seen in Fig 4b, there are a few defects located in between the nanodots. However, the volume fraction above the plane of the substrate occupied by these defects is small. A careful analysis of the AFM image of Fig. 4b (See Supporting Information) shows that the total volume occupied by all the defects represents less than 0.6% of the total volume occupied by the nanodots.

In conclusion, we demonstrated that the fabrication of metal nanodot arrays is feasible by using exclusively laser processing techniques. Laser fabricated porous alumina membranes consist of a thin alumina film which is drilled by near-field enhancement through a monolayer of $SiO_2$ spheres. The use of this mask allows for the localized and ordered deposition of small amount of material with a pre-defined geometry. The size of the dots thus generated is the same as that of the holes drilled on the PAM. This approach is quite general and can be applied to wide range of metals, semiconductors or complex oxides, especially if PLD is used for the material deposition step. This technique can be extended to spheres of variable diameter, thus allowing the distance between the nanodots to be varied for a defined application. However, Mie theory calculations show that the improvement in resolution (size of the near-field enhancement region) induced by the reduction of the sphere size will not be dramatic.[35] Focusing at a scale of a few nanometers is still a challenge that simple particle-masks will not solve. To improve resolution and accuracy, one of the possible extensions of the technique will be to exploit the deterministic and non-linear character of the damage processes involved in femtosecond laser machining.[36]

Nanodot arrays on surfaces have numerous properties that allow for a range of new surface functionalities. For example, metal dots may serve as catalysts for nanowire growth[37] or they may be used to attach biomolecules.[38] Moreover, the use of the present method allows preparing plasmonic structures if metal dot arrays are formed on dielectric substrates. In microscopy for instance, it was demonstrated recently that these nano-periodic substrates would make possible to obtain wide-field images with near-field resolution without scanning a probe in the vicinity of the sample.[39-41]

### *Experimental*

(100) oriented silicon wafers (p-type, resistivity of ca. 5-10 $\Omega$cm) were used in the experiments. The thin alumina layer was prepared by pulsed laser deposition in vacuum, as described in the paper of Gonzalo et *al.*.[21] A smooth $Al_2O_3$ film was first deposited onto the Si substrate using a KrF ($\lambda_{las}$ = 248 nm) excimer laser beam (Lumonics, PM-800) operating at 10 Hz. The films were grown under vacuum at room temperature. The deposition rate was determined by means of a quartz crystal microbalance. The number of pulses was adjusted to achieve a 20 nm film thickness.

Subsequently, a commercial solution (Kisker-Biotech) of monodisperse silica spheres (50 mg/ml) with radii $R$ = 250 nm was spincoated onto the $Al_2O_3$ surface to obtain a monolayer of assembled spheres. The delivered colloidal solution was diluted (1:5) in isopropanol (IPA). The monolayer pattern was obtained after deposition of droplets of the solution on the center of the substrate spinning at appropriate speed.

The resulting "spheres-$Al_2O_3$-Si" stacks were illuminated by single light pulses provided by an ArF ($\lambda_{las}$ = 193 nm) laser source (Lambda Physik, LPX220i). The laser pulse duration was $\cong$ 15 ns. The light was incident normally on the substrate. The image relay technique permits a near-uniform irradiation of the target materials on a spot size of $\cong$ 2×2 mm$^2$. The laser-pulse energy was varied with the aid of a manually operated beam attenuator (Optec, AT4030). After irradiation, the samples were ultrasonicated in IPA at room temperature for 10 min, to remove residual particles and contaminations from the substrate.

A thin layer of gold (20 nm) was deposited by PLD through the LF-PAM. In order to favor the diffusion of the species impinging on the PAM, their kinetic energy ($E_k$) was adjusted at ~ 120 eV atom$^{-1}$. $E_k$ values were previously determined from time-of-flight emission spectroscopy measurements. Subsequent removal of the PAM matrix with NaOH (1M) led to the release of ordered Au nanodots on Si surface.

LF-PAM and resulting nanodot arrays were analysed using a field-emission scanning electron microscope (FESEM, Hitachi S-5000) and an atomic force microscope (AFM, Digital Instruments Nanoscope III).

**Figures**

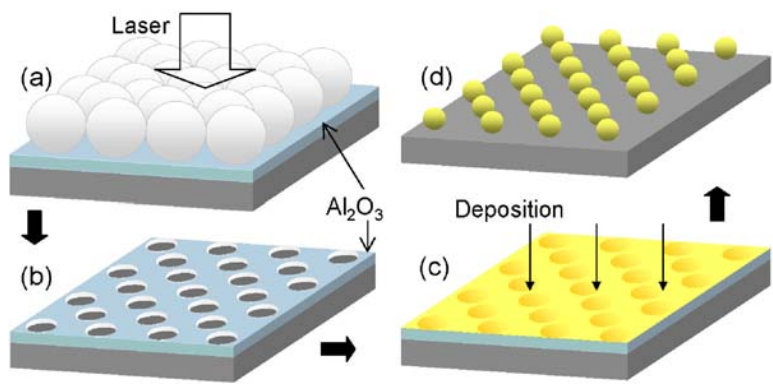

**Figure 1**    (color online) Nanodot array fabrication method: (a) A monolayer of spheres deposited on a thin alumina film is illuminated with a single nanosecond laser pulse. (b) Near-Field enhancement underneath the spheres leads to the parallel nanodrilling of the film. (c) A metal (gold in our case) is then deposited and the alumina membrane is dissolved in basic solution. (d) An ordered gold nanodot array onto the silicon substrate is then obtained.

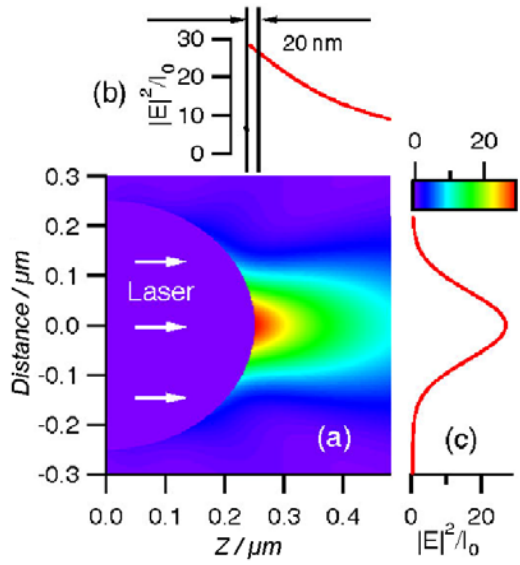

**Figure 2**   (color online) Focusing power of a 250-nm radius SiO$_2$ sphere: (a) Calculated optical near field distribution $|E|^2$ in the vicinity of the sphere ($n$=1.6) irradiated by a plane wave ($\lambda_{las}$=193 nm) incident from the left.  Variation of the external enhancement factor (b) along the propagation axis ($x$=0) and (c) on the substrate surface ($z$=250 nm).

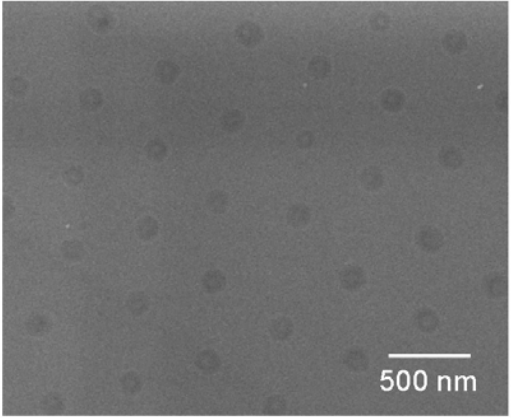

**Figure 3**     SEM micrograph of an $Al_2O_3$ thin film (20 nm) deposited on a Si substrate and simultaneously drilled by the near-field enhancement of a single nanosecond laser pulse which is produced by a lattice of $SiO_2$ spheres ($R$=250 nm).

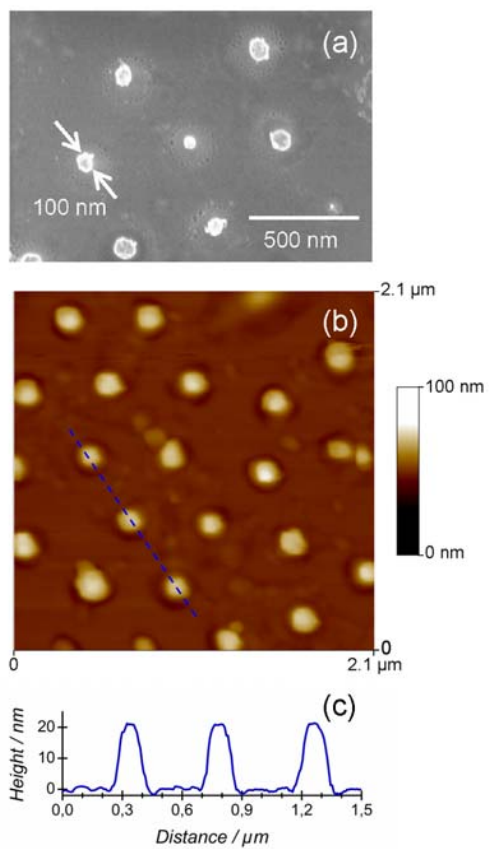

**Figure 4**    (color online) SEM image (a), AFM image (b) and depth profile (c) of gold nanodots created on silicon substrates by the LF-PAM based process.